\documentclass [10pt]{article}
\title{Physical Components, Coordinate Components,\\and the Speed of Light}
\author{Robert D. Klauber\\1100 University Manor Dr., 38B, Fairfield, IA 52556, USA\\rklauber@netscape.net}
\date{May 15, 2001}
\usepackage {graphicx}
\usepackage {longtable}
\begin{document}
\maketitle
\begin{abstract}

For generalized coordinate systems, the numerical values of vector and 
tensor components do not generally equal the physical values, i.e., the 
values one would measure with standard physical instruments. Hence, 
calculating physical components from coordinate components is important for 
comparing experiment with theory. Surprisingly, however, this calculational 
method is not widely known among physicists, and is rarely taught in 
relativity courses, though it is commonly employed in at least one other 
field (applied mechanics.) Different derivations of this method, ranging 
from elementary to advanced level, are presented. The result is then applied 
to clarify the oftentimes confusing issue of whether or not the speed of 
light in non-inertial frames is equal to $c$.

\end{abstract}

\section{INTRODUCTION}
\label{sec:introductionzed}

Although orthonormal (Cartesian and Lorentz) coordinate systems are the 
easiest to understand, they are not always the easiest to use to solve a 
given problem. An approximately infinite straight line wire antenna, for 
example, is far easier to analyze in a cylindrical coordinate system than a 
Cartesian one. More complicated coordinate systems (examples include 
spherical and toroidal systems) are also often used to make geometrically 
complex problems tractable. And in the most general case, one can imagine, 
as Einstein did with his coordinate ``mollusk'', a collection of grid lines 
spread throughout space in a somewhat haphazard manner, crossing one another 
in no particular pattern, and having widely varying degrees of spacing 
between lines.

In such non-orthonormal systems, called \textit{generalized coordinate} systems, the grid lines are 
numbered sequentially as 0,1,2,3,..., so one can always pinpoint one's 
location by specifying the number of the grid lines where one is located. 
Unlike orthonormal systems, however, these numbers do not directly reflect 
one's distance from the origin. If I am on a 2D flat surface at the 
intersection of generalized coordinate grid lines numbered 3 and 4, my 
coordinate location is (3,4), but I am not 3 physically measured meters (we 
will use meters as our measuring unit) from location (0,4). Depending on the 
spacing of the lines I could be 1 meter, 100 meters, a 1/2 meter, or any 
physical distance at all from (0,4). I can, of course, determine that 
distance experimentally by measuring it directly with a series of meter 
sticks laid end to end. But is there any way I can determine it analytically 
from my coordinate location values?

Before we begin our answer to that question, consider another situation 
where I wish to measure the speed of an object, but I can only count the 
number of grid lines the object passes by per second in my non-Cartesian 
system. How could I translate that speed into the speed I would measure 
using standard physical meter sticks instead of grid lines? The first speed 
(grid lines per second) is a \textit{coordinate speed}. It is relative to the coordinate system we 
use. Different spacing of the grid lines means a different number of lines 
passed per second, and that means a different coordinate speed. The second 
speed (meters per second = standard meter sticks passed per second) is 
unique, however. It is the same regardless of the amount of squeezing or 
spreading out of our gridlines. It is a \textit{physical speed}.

Each component of the velocity vector equals the speed in a given coordinate 
axis direction. If these components are in terms of grid lines per second, 
they are called \textit{coordinate components} of velocity. If they are in terms of meter sticks per 
second, they are called \textit{physical components.} For the special case where our coordinate grid is 
Cartesian (or Lorentzian), these components are identical, i.e., coordinate 
components equal the physical components we would measure.

Solutions to problems are typically found in terms of the coordinate system 
used, and hence are coordinate component solutions. But for non-orthonormal 
coordinates, we need to be able to relate those components to the values we 
would actually measure in experiment. In other words, we need to know how to 
find physical components from coordinate components.

Though widely used in applied mechanics\cite{References:1}$^{{\rm 
,}}$\cite{Lawrence:1969}$^{{\rm ,}}$\cite{Fung:1965}$^{{\rm 
,}}$\cite{Cemal:1962}$^{{\rm ,}}$\cite{Chung:1988}, the method for doing 
this is rarely taught in general relativity courses and seemingly known to 
but a small minority of physicists. In fact, the author knows of only a 
single relativity text (Misner, Thorne, and Wheeler\cite{Charles:1973}) 
where the subject is even mentioned. This seeming oversight is probably due, 
in some measure, to the fact that invariant quantities are often those 
measured in experiments. For example, the time dilation experienced by a 
particle in a cyclotron is the proper time \textit{$\tau $} for that particle, and this is 
the same (invariant) for any coordinate system (orthonormal or not) used to 
calculate it. However, there are situations in which vector or tensor 
components (rather than invariant quantities) are measured, and since these 
can be different from the theoretically determined coordinate components, we 
need to be able to deduce the former from the latter.

This article provides an introduction to physical components, explains how 
they can be calculated, and illustrates their utility with two examples. One 
of these provides an answer to a question many students are commonly 
perplexed by. That is, ``Is the speed of light in a non-inertial (general 
relativistic) frame equal to $c$?''

\section{PHYSICAL COMPONENTS: THE SIMPLEST VIEW}
\label{sec:physical}

Readers with marginal background in general relativity may wish to read the 
present more elementary and pedagogic section on physical components, and 
then skip to Section \ref{sec:mylabel1}. Sections 
\ref{sec:mylabel2} and \ref{sec:illustrating} use greater 
mathematical rigor to draw the same conclusions as Section 
\ref{sec:physical} and cover more advanced topics such as tensor 
analysis. 

We begin this section by refining the concept of generalized coordinates, 
i.e., those which may be other than Cartesian (or Lorentzian.) Note that 
generalized coordinates are by custom labeled with superscripts. That is, 
instead of three dimensions labeled x,y, and z, or x$_{{\rm 1}{\rm ,}{\rm 
}}$x$_{{\rm 2}}$, and x$_{{\rm 3}}$, we use x$^{{\rm 1}}$, x$^{{\rm 2}}$, 
and x$^{{\rm 3}}$. Raising to a power is indicated using parenthesis, i.e., 
(x$^{{\rm 1}}$)$^{{\rm 2}}$. 

To avoid confusion we deal solely with flat spaces until noted otherwise.

\subsection{Generalized Coordinates}
\label{subsec:generalized}

Consider the 2D flat space (no time dimension for the present) of Figure 1. 
Grid lines are numbered sequentially with integers. If we use a Cartesian 
coordinate system with $X^{{\rm 1}}$ and $X^{{\rm 2}}$ grid lines spaced one 
standard meter stick apart, then the distance \textit{ds} between any two 
(infinitesimally close) points is found from the Pythagorean theorem

\begin{equation}
\label{eq1}
(ds)^{2} = (dX^{1})^{2} + (dX^{2})^{2}
\end{equation}

\noindent
where \textit{dX}$^{{\rm 1}}$ and \textit{dX}$^{{\rm 2}{\rm} }$ values equal physical (standard) 
meter stick distances in orthogonal directions. The reader should not be 
confused by the somewhat loose identification of the infinitesimal ``$d$'' in 
this discussion with the finite difference ``$\Delta $'' in the finite size 
figures.

\begin{figure}
\centering
\includegraphics[bbllx=0.26in,bblly=0.13in,bburx=6.75in,bbury=2.83in,scale=1.00]{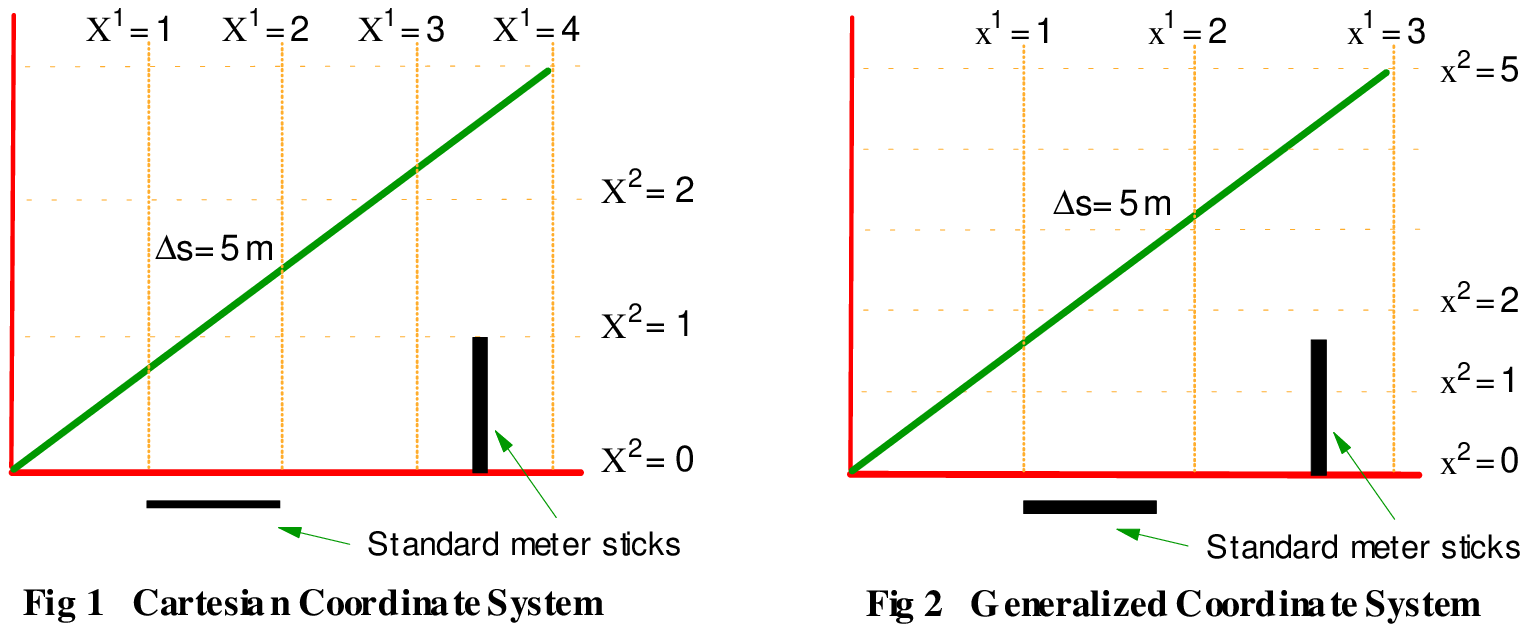}
\end{figure}

\bigskip

If, as in Figure 2, we change the grid line spacing to form a different, 
non-Cartesian, coordinate grid ($x^{{\rm 1}}$,$ x^{{\rm 2}}$) for the same flat 
space, then distances between successive $x^{{\rm 1}}$ and $x^{{\rm 2}}$ grid 
lines are no longer one meter stick. So in the new grid, successive lines 
are still labeled as $x^{{\rm 1}}$=1, $x^{{\rm 1}}$=2, $x^{{\rm 1}}$=3, etc, 
but the distances between successive $x^{{\rm 1}}$ (and $x^{{\rm 2}}$) grid 
lines are no longer one meter. For example, the distance between the 
$x^{{\rm 1}}$=1 and $x^{{\rm 1}}$=2 lines is not one meter, even though 
$\Delta x^{{\rm 1}}$=1 between those lines. So \textit{dx}$^{{\rm 1}{\rm} }$and 
\textit{dx}$^{{\rm 2}}$ values equal number of grid lines between points, \textit{not} physical 
meter stick distances (i.e., not number of meters) between the points, and 
the form of (\ref{eq1}) for the new coordinates no longer holds. That is, for \textit{ds} equal 
to the distance in meters between the same two points,

\begin{equation}
\label{eq2}
(ds)^{2} \ne (dx^{1})^{2} + (dx^{2})^{2}{\rm .}
\end{equation}

We therefore modify (\ref{eq1}) to

\begin{equation}
\label{eq3}
(ds)^{2} = g_{11} (dx^{1})^{2} + g_{22} (dx^{2})^{2}
\end{equation}

\noindent
where g$_{{\rm 1}{\rm 1}}$ and g$_{{\rm 2}{\rm 2}}$ must be chosen such that 
(\ref{eq1}) remains valid, i.e., $g_{{\rm 1}{\rm 1}}$ and $g_{{\rm 2}{\rm 2}}$ satisfy

\begin{equation}
\label{eq4}
g_{11} (dx^{1})^{2} = (dX^{1})^{2}\quad \quad g_{22} (dx^{2})^{2} = 
(dX^{2})^{2}{\rm ,}
\end{equation}

\noindent
or more simply,

\begin{equation}
\label{eq5}
dX^{1} = \sqrt {g_{11}}  dx^{1}\quad \quad \quad dX^{2} = \sqrt {g_{22}}  
dx^{2}{\rm .}
\end{equation}

In both (\ref{eq1}) and (\ref{eq3}) \textit{ds} is the same distance (in meter stick lengths) between 
the same two points (\textit{ds} is invariant between coordinate systems), and each 
expression is called the \textit{line element} between the two points for that particular 
coordinate grid. The term generalized coordinates is used for such 
non-Cartesian coordinates because the coordinate grid ($x^{{\rm 
1}}$,$x^{{\rm 2}}$) is completely arbitrary (general). The quantity 
$g_{{\rm i}{\rm j}}$ ($i,j $= 1,2) is known as the \textit{metric} of the coordinate grid. Note 
that for different generalized coordinate grids ($x^{{\rm 1}}$,$x^{{\rm 2}}$), 
there will be different values for $g_{{\rm i}{\rm j}}$.

\subsection{Physical vs. Coordinate Components}
\label{subsec:physical}

\textit{dX}$^{i}$ in (\ref{eq5}) are physically measured values (i.e., they are the number of 
meter sticks one would find physically by measuring in the $X^{{\rm 1}}$ and 
$X^{{\rm 2}}$ directions), and hence they are unique. The quantities 
\textit{dx}$^{i}$, on the other hand, represent number of grid lines one would count 
off in the ($x^{{\rm 1}}$,$x^{{\rm 2}}$) system, and have no physical meaning, 
since they are different for each arbitrary coordinate grid. Hence, 
\textit{dX}$^{i}$ are called \textit{physical components}\cite{References:1}$^{{\rm ,}}$\cite{Charles:1973}; 
and \textit{dx}$^{i}$, \textit{coordinate components.} 

The same relationship [see (\ref{eq5})] between components in Cartesian and 
generalized systems exists for vectors other than \textit{dx}$^{{\rm i}}$. For any 
vector expressed as $v^{{\rm i}{\rm} }$in a generalized coordinate system, we 
can find the equivalent physical component values $v^{\hat {i}}$ an 
experimentalist would actually measure via

\begin{equation}
\label{eq6}
v^{\hat {i}} = \sqrt {g_{\underline {i} \underline {i}} }  v^{i}{\rm ,}
\end{equation}

\noindent
where underlining implies no summation, and we have introduced the notation 
used by Misner, Thorne, and Wheeler\cite{Charles:1973} of carets over 
component indices to designate physical values.

\subsection{Non-orthogonal Coordinate Grids}
\label{subsec:mylabel1}

Note that if the coordinate axes in (\ref{eq1}) are not orthogonal, then (\ref{eq1}) and 
hence (\ref{eq3}), get an extra cross term, i.e.,

\begin{equation}
\label{eq7}
(ds)^{2} = (dX^{1})^{2} + (dX^{2})^{2} + 2dX^{1}dX^{2}\cos \theta = g_{ij} 
dx^{i}dx^{j}
\end{equation}

\noindent
where repeated indices imply summation, \textit{$\theta $}  is the angle between the $X^{{\rm 
1}}$ and $X^{{\rm 2}}$ axes, and $g_{12} \ne 0$.

\subsection{Non-flat Spaces}
\label{subsec:mylabel2}

A small enough area on a curved surface such as that of a sphere appears 
approximately flat. Since we dealt only with differential lengths in 
relations (\ref{eq1}) through (\ref{eq7}), all of those relations remain valid (locally) on 
a curved surface. In particular, we find physical components (those measured 
experimentally) in the same manner, i.e., with (\ref{eq6}), for any vector quantity 
in either a flat or non-flat space. 

\subsection{Spacetime}
\label{subsec:spacetimenough}

In special relativity, the four dimensional analogue of the Pythagorean 
theorem (\ref{eq1}) is

\begin{equation}
\label{eq8}
(ds)^{2} = - (cdT)^{2} + (dX^{1})^{2} + (dX^{2})^{2} + (dX^{3})^{2}
\end{equation}

\noindent
where $c$ is the speed of light, $T$ is physical time, and \textit{ds} is now the \textit{spacetime interval,} an amalgam 
of both the physical spatial length and the physical time between two 4D 
\textit{events}. Note that special relativity deals primarily with physical components and 
orthogonal grid lines in time and space, as in (\ref{eq8}). If \textit{ds} = 0, then the 
physical length in (\ref{eq8}) (e.g., \textit{dX}$^{{\rm 1}}$ where \textit{dX}$^{{\rm 2}}$ = \textit{dX}$^{{\rm 3}}$ 
= 0) divided by the physical time \textit{dT} yields $c$, the physical speed of light. 
This result can be generalized, so if light travels between two events, then 
we always have \textit{ds} =0 between those events. (Because of this, light is often 
said to travel a \textit{null} path.)

In general relativity, one uses generalized 4D coordinate systems, and the 
analogue to (\ref{eq3}) is

\begin{equation}
\label{eq9}
ds^{2} = g_{00} (dx^{0})^{2} + g_{11} (dx^{1})^{2} + g_{22} (dx^{2})^{2} + 
g_{33} (dx^{3})^{2} = g_{\mu \nu}  dx^{\mu} dx^{\nu} 
\end{equation}

\noindent
where \textit{dx}$^{{\rm 0}}$ = \textit{cdt}, $g_{00} < 0$, Greek indices are used to designate 
four dimensional (relativistic) quantities, and repeated indices imply 
summation over those indices.$^{{\rm} }$\textit{} Here, \textit{dt} represents \textit{coordinate time}, which is an 
arbitrarily defined time (just as our spatial coordinate grids were 
arbitrarily defined) generally different from the physical time one would 
measure experimentally with standard clocks. 

Use of coordinate time can simplify the solution of many problems. It can be 
visualized in a 3D connotation as a set of clocks that run at rates 
different than standard (physical) clocks and that are distributed at every 
point in our generalized 3D coordinate grid. In a 4D connotation, coordinate 
time can be visualized as a set of grid lines whose labels increment 
sequentially in the time direction, but which do not correspond to one 
second between grid lines. 

Physical time is found from coordinate time in analogous fashion to (\ref{eq5}) or 
(\ref{eq6}), i.e.,

\begin{equation}
\label{eq10}
{\rm p}{\rm h}{\rm y}{\rm s}{\rm i}{\rm c}{\rm a}{\rm l}\;{\rm t}{\rm i}{\rm 
m}{\rm e}\;{\rm i}{\rm n}{\rm t}{\rm e}{\rm r}{\rm v}{\rm a}{\rm l}{\rm 
}{\rm =} {\rm} d\hat {t} = \sqrt { - g_{00}}  dt;\quad \quad \quad dx^{\hat 
{0}} = \sqrt { - g_{00}}  dx^{0}{\rm .}
\end{equation}

Note that by substituting the coordinate component values of (\ref{eq5}) 
[incorporating the third spatial dimension] and (\ref{eq10}) into (\ref{eq9}), we get (\ref{eq8}), 
where the components in (\ref{eq8}) are physical components for an observer at rest 
in the generalized grid system. Since (\ref{eq8}) represents a special relativistic, 
or Lorentz, coordinate system we can conclude the following.

\textit{In general relativity, for spacetime reference frames whose line elements can be cast in the form of} (\ref{eq9})\textit{, local physical components at any 4D point (event) are equal to those in a local} \textit{co-moving}\textit{ Lorentz frame at the same 4D point.}\cite{This:1}

Note further that for non-orthogonal axes, cross terms would be added to the 
line element (\ref{eq9}) analogously to (\ref{eq7}). If such cross terms are between time 
and space [e.g., a non-zero $g_{02}$\textit{(cdt)(dx}$^{{\rm 2}}$)\textit{} term would be present in 
(\ref{eq9})] then time is not orthogonal to space\cite{Such:1} and, it can be shown 
that in such frames the above conclusion is not 
valid\cite{Robert:1998}$^{{\rm ,}}$\cite{Robert:1}.

\section{PHYSICAL COMPONENTS: THE BASIS VECTOR VIEW}
\label{sec:mylabel2}

This and the following section are intended for readers having a reasonable 
foundation in general relativity. In particular, such readers should already 
be familiar with the concept of basis vectors and the metric identity

\begin{equation}
\label{eq11}
g_{ij} = {\rm {\bf e}}_{i} \cdot {\rm {\bf e}}_{j} {\rm ,}
\end{equation}

\noindent
where \textbf{e}$_{{\rm i}}$ is a generalized coordinate basis vector.

\subsection{Unit and Generalized Basis Vectors}
\label{subsec:mylabel3}

The displacement vector $d$\textbf{x} between two points in a 2D Cartesian 
coordinate system is

\begin{equation}
\label{eq12}
d{\rm {\bf x}}\, = dX^{1}{\rm {\bf \hat {e}}}_{1} + dX^{2}{\rm {\bf \hat 
{e}}}_{2} 
\end{equation}

\noindent
where the ${\rm {\bf \hat {e}}}_{i} $ are unit basis vectors and \textit{dX}$^{{\rm 
i}}$ are physical components. For the same vector $d$\textbf{x} expressed in a 
different, generalized, coordinate system we have different coordinate 
components \textit{dx}$^{{\rm i}{\rm} } \ne $ \textit{dX}$^{{\rm} {\rm i}}$, but a similar 
expression

\begin{equation}
\label{eq13}
d{\rm {\bf x}}\, = dx^{1}{\rm {\bf e}}_{1} + dx^{2}{\rm {\bf e}}_{2} {\rm 
,}
\end{equation}

\noindent
where the generalized basis vectors \textbf{e}$_{{\rm i}{\rm} {\rm} }$point 
in the same directions as the corresponding unit basis vectors ${\rm {\bf 
\hat {e}}}_{i} $, but are not equal to them. Hence, for ${\rm {\bf \hat 
{e}}}_{1} $, we have

\begin{equation}
\label{eq14}
{\rm {\bf \hat {e}}}_{1} = {\frac{{{\rm {\bf e}}_{1}} }{{\vert {\rm {\bf 
e}}_{1} \vert} }} = {\frac{{{\rm {\bf e}}_{1}} }{{\sqrt {{\rm {\bf e}}_{1} 
\cdot {\rm {\bf e}}_{1}} } }} = {\frac{{{\rm {\bf e}}_{1}} }{{\sqrt {g_{11} 
}} }}
\end{equation}

\noindent
where we used (\ref{eq11}) to get the RHS. In general, we have

\begin{equation}
\label{eq15}
{\rm {\bf \hat {e}}}_{i} = {\frac{{{\rm {\bf e}}_{i}} }{{\vert {\rm {\bf 
e}}_{i} \vert} }} = {\frac{{{\rm {\bf e}}_{i}} }{{\sqrt {{\rm {\bf 
e}}_{\underline {i}}  \cdot {\rm {\bf e}}_{\underline {i}} } } }} = 
{\frac{{{\rm {\bf e}}_{i}} }{{\sqrt {g_{\underline {i} \underline {i}} }  
}}}
\end{equation}

\noindent
where again, underlining implies no summation.

Substituting (\ref{eq15}) into (\ref{eq12}) and equating with (\ref{eq13}), one obtains

\begin{equation}
\label{eq16}
dX^{1} = \sqrt {g_{11}}  dx^{1}\quad \quad \quad dX^{2} = \sqrt {g_{22}}  
dx^{2}{\rm ,}
\end{equation}

\noindent
which is the same relationship between displacement physical and coordinate 
components as (\ref{eq5}).

Consider a more general case of an arbitrary vector \textbf{v}

\begin{equation}
\label{eq17}
{\rm {\bf v}} = v^{1}{\rm {\bf e}}_{1} + v^{2}{\rm {\bf e}}_{2} = v^{\hat 
{1}}{\rm {\bf \hat {e}}}_{1} + v^{\hat {2}}{\rm {\bf \hat {e}}}_{2} 
\end{equation}

\noindent
where, \textbf{e}$_{{\rm 1}}$ and \textbf{e}$_{{\rm 2}}$ here do not, in 
general, have to be orthogonal, \textbf{e}$_{i}$ and ${\rm {\bf \hat 
{e}}}_{i} $ point in the same direction for each index $i$, and as before, 
carets over component indices indicate physical components. Substituting 
(\ref{eq15}) into (\ref{eq17}), one readily obtains

\begin{equation}
\label{eq18}
v^{\hat {i}} = \sqrt {g_{\underline {i} \underline {i}} }  v^{i}{\rm ,}
\end{equation}

\noindent
which is the same as (\ref{eq6}), and which we have shown here to be valid in both 
orthogonal and non-orthogonal systems.

In consonance with the above, many authors\cite{For:1} note that the 
physical component of any vector in a given direction is merely the 
projection of that vector onto that direction, i.e., the inner product of 
the vector with a unit vector in the given direction. This is simply the 
component of the vector in a basis having a unit basis vector pointing in 
that direction.

If we apply this prescription by taking the dot product of ${\rm {\bf \hat 
{e}}}_{1} $with \textbf{v} of (\ref{eq17}), then use (\ref{eq15}) and (\ref{eq11}), we end up with 
(\ref{eq18}) for $i$=1. This is obviously generalized to any index $i$.

As a further aid to those readers familiar with anholonomic coordinates 
(which superimpose unit basis vectors on a generalized coordinate grid), we 
reference Eringen's\cite{Cemal:1971} comment ``.. the anholonomic 
components .. of a ... vector ... are identical to [physical components]''.

Following similar logic to that used in Section \ref{sec:physical}, 
one sees (\ref{eq18}) is valid locally in curved, as well as flat, spaces, and can 
be extrapolated to 4D general relativistic applications. So, very generally, 
for a 4D vector $u^{{\rm \mu} }$

\begin{equation}
\label{eq19}
u^{\hat {i}} = \sqrt {g_{\underline {i} \underline {i}} }  u^{i}\quad \quad 
u^{\hat {0}} = \sqrt { - g_{00}}  u^{0}{\rm ,}
\end{equation}

\noindent
where Roman sub and superscripts refer solely to spatial components (i.e. 
$i$ = 1,2,3.)

\subsection{Contravariant vs. Covariant Components}
\label{subsec:contravariant}

As the reader of this section is no doubt aware, coordinate displacement 
vector components \textit{dx}$^{i} $(or \textit{dx}$^{\mu} $in relativity theory) are 
\textit{contravariant} components, as are the components of the vector expressed in (\ref{eq19}). A similar 
derivation to that shown above carried out for \textit{covariant} components leads to the 
following relationship between covariant coordinate components and covariant 
physical components

\begin{equation}
\label{eq20}
u_{\hat {i}} = \sqrt {g^{\underline {i} \underline {i}} } u_{i} \quad \quad 
u_{\hat {0}} = \sqrt { - g^{00}} u_{0} {\rm .}
\end{equation}

\subsection{Tensors}

Considering second order tensors as open, or dyadic, products of vectors, 
with each tensor component having two basis vectors associated with it, one 
can use the general methodology of Sub-section 
\ref{subsec:mylabel3} to derive the following relationships

\begin{equation}
\label{eq21}
T^{\hat {\mu} \hat {\upsilon} } = \sqrt {g_{\underline {\mu}  \underline 
{\mu} } }  \sqrt {g_{\underline {\upsilon}  \underline {\upsilon} } }  
T^{\mu \upsilon} \quad \quad T_{\hat {\mu} \hat {\upsilon} } = \sqrt 
{g^{\underline {\mu}  \underline {\mu} } } \sqrt {g^{\underline {\upsilon}  
\underline {\upsilon} } } T_{\mu \upsilon}  {\rm .}
\end{equation}

As a first sample application of physical components, consider the 4D 
electromagnetic tensor $F^{{\rm \mu} {\rm \nu} }$ containing as components 
the electric and magnetic induction fields \textbf{E} and 
\textbf{B}\cite{See:1}$^{{\rm ,}}$\cite{John:1975}\textbf{.}  Assume a 
cylindrical coordinate system ($t,r,\phi $,z) with a field in the axial $z$ 
direction measured to be $B_{m}$ (subscript for ``measured'') and with no 
other electric or magnetic field components. The physical (measured) 
component of the magnetic induction in the $z$ direction is $B_{m}$, and the 
corresponding term in the $F^{{\rm \mu} {\rm \nu} }$ tensor is the $\mu $ = 
$r$, $\nu $ = $\phi $ component. Hence,

\begin{equation}
\label{eq22}
F^{\hat {r}\hat {\phi} } = B_{m} {\rm .}
\end{equation}

The metric for a cylindrical coordinate system has $g_{{\rm r}{\rm r}{\rm 
}}$= 1 and $g_{{\rm \phi} {\rm \phi} }$ = $r^{{\rm 2}}$. Using these with\textit{} (\ref{eq21}), 
we then find the generalized tensor component, which should be used in any 
tensor analysis of the problem in cylindrical coordinates, to be

\begin{equation}
\label{eq23}
F^{r\phi}  = {\frac{{F^{\hat {r}\hat {\phi} }}}{{\sqrt {g_{rr}}  \sqrt 
{g_{\phi \phi} } } }} = {\frac{{B_{m}} }{{r}}}{\rm .}
\end{equation}

We note that tensor analysis must be carried out with coordinate components 
($F^{\mu \nu} $ here), not physical components (i.e., not $F^{\hat {\mu} \hat 
{\upsilon} }$.) Fung\cite{See:2} observes that ``... physical components 
... do not transform according to the tensor transformation laws and are not 
components of tensors.''

Hence, at the beginning of an analysis we must first convert known measured 
(physical) component values to coordinate component values. Then we carry 
out the vector/tensor analysis. As a final step, in order to compare our 
results with experiment, we convert the coordinate component answer to 
physical component form.

\section{ILLUSTRATING THE MOST GENERAL CASE}
\label{sec:illustrating}

\subsection{Visualizing basis vectors and physical components}

The import of physical and coordinate values for both contravariant and 
covariant components can be visualized graphically. We start with two purely 
spatial sets of coordinate grid lines ($x^{{\rm 1}}$,$x^{{\rm 2}}$), which are 
quite arbitrary, and in this case, are not orthogonal. The contravariant 
component representation of a certain vector \textbf{V} is shown in Figure 
3; the covariant representation for the same vector in Figure 4. Vector 
\textbf{V} is located at $x^{{\rm 1}}$ = 2 grid units, $x^{{\rm 2}}$ = 10 grid 
units and has magnitude and direction as shown. 

Covariant basis vectors \textbf{e}$_{{\rm i}}$ (Figure 3) for use with 
contravariant components are defined as, for example,

\begin{equation}
\label{eq24}
{\rm {\bf e}}_{2} = {\left. {{\frac{{\partial \,P}}{{\partial x^{2}}}}} 
\right|}_{x^{1} = 2} = {\frac{{{\rm 2}\;{\rm m}{\rm e}{\rm t}{\rm e}{\rm 
r}\;{\rm s}{\rm t}{\rm i}{\rm c}{\rm k}{\rm s}}}{{{\rm 1}\;{\rm x}^{{\rm 
2}}\;{\rm g}{\rm r}{\rm i}{\rm d}\;{\rm u}{\rm n}{\rm i}{\rm t}}}} = 2{\rm 
{\bf \hat {e}}}_{2} 
\end{equation}

\noindent
where ${\rm {\bf \hat {e}}}_{2} $is the unit vector in the direction of 
\textbf{e}$_{{\rm 2}}$. Similar logic holds for \textbf{e}$_{{\rm 1}{\rm 
,}{\rm} }$and we find in this example that \textbf{e}$_{{\rm 1}}$ = ${\rm 
{\bf \hat {e}}}_{1} $. The coordinate component of \textbf{V} in the 
\textbf{e}$_{{\rm 2}}$ direction is $V^{2} = 0$, the number of 
\textbf{e}$_{{\rm 2}}$ basis vectors in that direction. The physical 
component in the same direction is $V^{\hat {2}} = 4$, the number of unit 
basis vectors ${\rm {\bf \hat {e}}}\,_{2} $in that direction. This physical 
component is the same regardless of what grid we choose, whereas the 
coordinate component varies with the grid spacing.

\begin{equation}
\label{eq25}
V^{\hat {2}} = \;\vert {\rm {\bf e}}_{2} \vert V^{2} = \sqrt {{\rm {\bf 
e}}_{2} \cdot {\rm {\bf e}}_{2}}  \,V^{2} = \sqrt {g_{22}}  V^{2} = \sqrt {2 
\cdot 2} \,2 = 4
\end{equation}

The physical component is the numerical value associated with the unit 
length basis vector ${\rm {\bf \hat {e}}}_{2} $in the \textbf{e}$_{{\rm 2}}$ 
direction.

As an aside, $g_{ij} = {\rm {\bf e}}_{i} \cdot {\rm {\bf e}}_{j} $ is 
obviously not diagonal, as ${\rm {\bf e}}_{i} \cdot {\rm {\bf e}}_{j} \ne 
0$.

Contravariant basis vectors \textbf{e}$^{{\rm j}}$ (called \textit{one-forms} typically in 
general relativity) are defined, with a purpose in mind, by the relation

\begin{equation}
\label{eq26}
{\rm {\bf e}}_{i} \cdot {\rm {\bf e}}^{j} = \delta _{i} ^{j}{\rm .}
\end{equation}

With (\ref{eq26}) we get the contravariant basis vectors (or one-forms) shown in 
Figure 4. The covariant components $V_{1}$ and $V_{2}$ represent the number of 
contravariant (one-form) basis vectors \textbf{e}$^{{\rm 1}}$ and 
\textbf{e}$^{{\rm 2}}$, respectively, it takes along the \textbf{e}$^{{\rm 
1}}$ and \textbf{e}$^{{\rm 2}}$ directions, respectively, to give us vector 
components along those directions that will vector sum to yield \textbf{V}. 
Of course, \textbf{e}$^{{\rm 2}}$ and \textbf{e}$_{{\rm 2}}$ are not aligned 
(as they would be for orthogonal grid lines). Neither are \textbf{e}$^{{\rm 
1}}$ and \textbf{e}$_{{\rm 1}}$ aligned.

\begin{figure}
\centering
\includegraphics[bbllx=0.26in,bblly=0.13in,bburx=7.24in,bbury=2.74in,scale=1.00]{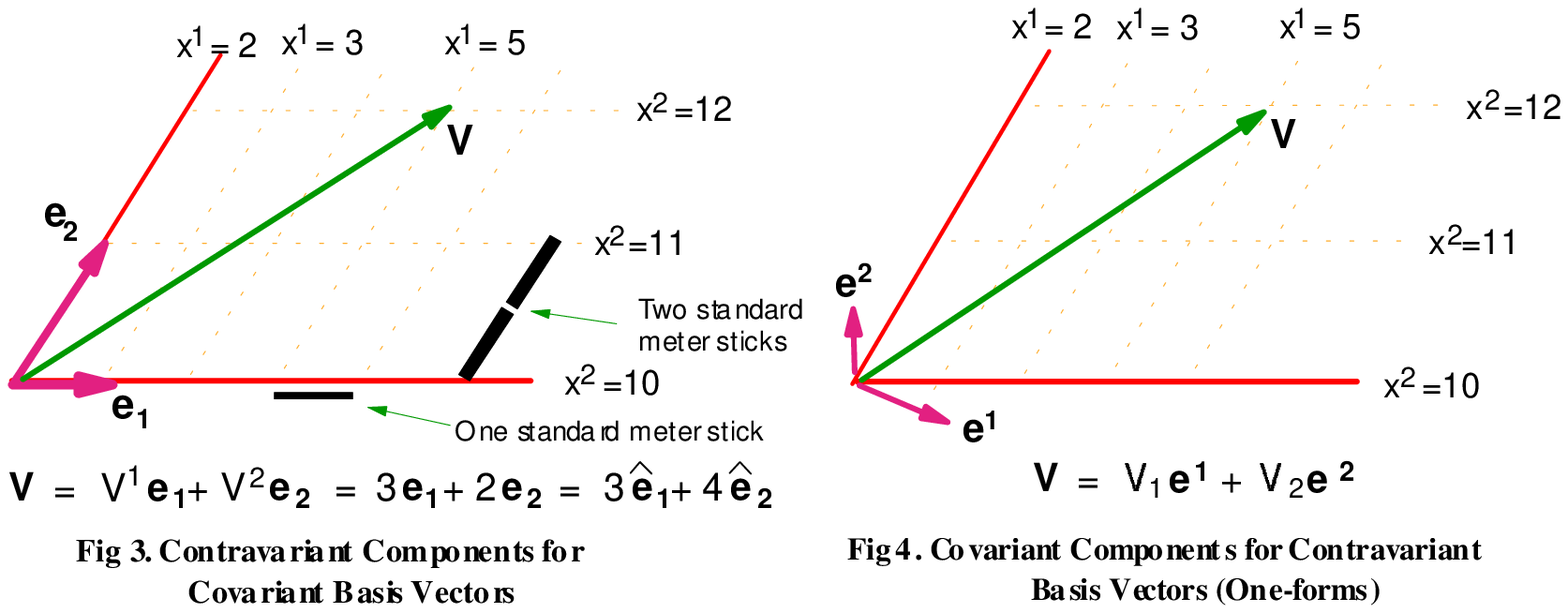}
\end{figure}

\bigskip

As with the contravariant component case, we can get determine any physical 
component, i.e., the number of unit vectors it will take along a given 
contravariant (one-form) basis vector direction for the vector component in 
that direction.

\begin{equation}
\label{eq27}
V_{\hat {2}} = \,\vert {\rm {\bf e}}^{2}\vert V_{2} = \sqrt {{\rm {\bf 
e}}^{2} \cdot {\rm {\bf e}}^{2}} \,V_{2} = \sqrt {g^{22}} \,V_{2} 
\end{equation}

Note the contravariant physical component $V^{\hat {2}}$ of (\ref{eq25}) does not 
equal the covariant physical component $V_{\hat {2}} $ of the above. The 
number of unit vectors needed in the \textbf{e}$^{{\rm 2}}$ direction is not 
the same as that in the \textbf{e}$_{{\rm 2}}$ direction.

\subsection{Non-time-orthogonal Axes}
\label{subsec:mylabel4}

Consider now the interesting case of spacetime around a star or black hole 
possessing angular momentum in which time is not orthogonal to space, i.e. 
time-space cross terms exist in the metric. In this case, instead of being 
spatial, our \textbf{e}$_{{\rm 2}}$ direction in Figure 3 can be thought of 
as representing time. So wherever we have used a ``2'' sub or super script, 
now think of it as a ``0''. We shall refer to such a system as 
non-time-orthogonal (NTO). For this application we will also consider 
\textbf{V} as our (infinitesimal) displacement four-vector. That is 
$V^{{\rm \mu} }$ = \textit{dx}$^{{\rm \mu} }$ here.

Note that a clock fixed at $x^{{\rm 1}}$ = 2 travels a path in spacetime 
along the \textbf{e}$_{{\rm 0}}$ (our old \textbf{e}$_{{\rm 2}}$) direction, 
NOT along the \textbf{e}$^{{\rm 0}{\rm} }$(old \textbf{e}$^{{\rm 2}}$) 
direction.

Of course we can now get coordinate values for our $\mu $ = 0 contravariant 
and covariant components, and of course they will generally be different. 
However what about physical (standard) time? How many actual seconds 
(analogous to actual meter sticks) will a standard physical clock travel in 
spacetime if it is fixed in space at $x^{{\rm 1}}$= 2? It is the physical 
component of the ``0'' component of \textit{dx}$^{{\rm 0}}$ = \textit{cdt} (divided by $c$, of 
course.) That is, it is found from equation (\ref{eq25}) with sub/superscripts 2 $ 
\to $ 0. It is NOT the physical component of \textit{dx}$_{{\rm 0}}$ (again divided by 
$c$) of equation (\ref{eq27}), which gives a different numerical value.

Why can't we use either the contravariant or covariant physical component 
for standard (physically measured) time? Because \textbf{e}$^{{\rm 2}}$ and 
\textbf{e}$_{{\rm 2}}$ do not point in the same direction. Specifically the 
component $dx_{\hat {0}} $ does not represent the number of physical seconds 
along the $x^{{\rm 1}}$ = constant = 2 direction. It represents the physical 
component in the \textbf{e}$^{{\rm 2}}$ direction, which does not represent 
a fixed spatial location within the given frame, but is actually a 
combination of both space and time.

Both contravariant and covariant representations are equivalent in the sense 
that both represent the same vector. For NTO frames, however, they are NOT 
equivalent in terms of how much of that\textbf{} vector is physically 
spacelike and how much is physically timelike.

So to represent time and space as they are actually measured with fixed 
clocks and fixed measuring meter sticks in such a frame, we have to use 
physical components associated with \textit{dx}$^{{\rm \mu} }$, NOT those of 
\textit{dx}$_{{\rm \mu} }$. Of course, for Lorentz frames it doesn't matter. And as far 
as physical components go, neither does it matter for Schwarzchild 
coordinates or any other time orthogonal (TO) frame. That is because in 
orthogonal coordinates, contravariant and covariant \textit{physical} components are 
identical. For NTO frames, however, it definitely does matter whether one 
uses \textit{dx}$^{{\rm \mu} }$\textit{} or \textit{dx}$_{{\rm \mu} }$ ,\textit{}  and when dealing with such frames 
one must proceed with caution.

In NTO frames, contravariant components of displacement correspond to true 
space and time directions; covariant displacement components do not. 
Therefore, four-velocity is, in the strictest sense, a contravariant vector 
since it equals \textit{dx}$^{{\rm \mu} }$/d$\tau $.

\subsection{General Case Conclusion}

Although it is immaterial in Lorentz frames whether one uses contravariant 
or covariant components, and it is immaterial in any TO frame whether one 
uses contravariant or covariant \textit{physical} components, it is critically important in 
NTO frames to use contravariant 4-vectors for displacement and velocity.

It will not matter in any case for invariants. They are not components of 
vectors or tensors, and will come out the same in any coordinates, in any 
kind of frame. However, for experimentally measured quantities that are 
components of a vector or tensor in an NTO frame, one must use physical 
components of the correct covariant or contravariant form. 

Note that while physical values are not invariant between frames, within a 
given frame, they are unique. They equal what an observer in that frame 
would measure.

\section{THE SPEED OF LIGHT}
\label{sec:mylabel1}

With an understanding of the difference between physical and coordinate 
components, one can now unravel the widespread confusion (in the author's 
experience at least) over the value of the speed of light in a gravitational 
field.

In a non-inertial (general relativistic) frame, the coordinate speed of 
light is found (see subsection \ref{subsec:spacetimenough}) by setting 
\textit{ds} = 0 in (\ref{eq9}) and solving for the length in generalized coordinates (the 
coordinate length = number of spatial grid lines) that the light ray travels 
divided by the time in generalized coordinates (the coordinate time = number 
of temporal grid lines) it takes to travel, i.e., for (\ref{eq9}) with \textit{dx}$^{{\rm 2}}$ 
= \textit{dx}$^{{\rm 3}}$ = 0,

\begin{equation}
\label{eq28}
{\frac{{dx^{1}}}{{dt}}} = \sqrt {{\frac{{ - g_{00}} }{{g_{11}} }}} \;c = 
{\frac{{{\rm (}{\rm n}{\rm u}{\rm m}{\rm b}{\rm e}{\rm r}\;{\rm o}{\rm 
f}\;{\rm s}{\rm p}{\rm a}{\rm t}{\rm i}{\rm a}{\rm l}\;{\rm g}{\rm r}{\rm 
i}{\rm d}\;{\rm l}{\rm i}{\rm n}{\rm e}{\rm s}\;{\rm t}{\rm r}{\rm a}{\rm 
v}{\rm e}{\rm r}{\rm s}{\rm e}{\rm d}{\rm )}}}{{{\rm (}{\rm n}{\rm u}{\rm 
m}{\rm b}{\rm e}{\rm r}\;{\rm o}{\rm f}\;{\rm c}{\rm o}{\rm o}{\rm r}{\rm 
d}{\rm i}{\rm n}{\rm a}{\rm t}{\rm e}\;{\rm t}{\rm i}{\rm m}{\rm e}\;{\rm 
u}{\rm n}{\rm i}{\rm t}{\rm s}\;{\rm p}{\rm a}{\rm s}{\rm s}{\rm e}{\rm 
d}{\rm )}}}}
\end{equation}

On the other hand, the physical speed of light is the physical length 
divided by the physical time,

\begin{equation}
\label{eq29}
{\frac{{\sqrt {g_{11}}  dx^{1}}}{{\sqrt { - g_{00}}  dt}}} = \;c = 
{\frac{{{\rm (}{\rm n}{\rm u}{\rm m}{\rm b}{\rm e}{\rm r}\;{\rm o}{\rm 
f}\;{\rm s}{\rm t}{\rm a}{\rm n}{\rm d}{\rm a}{\rm r}{\rm d}\;{\rm m}{\rm 
e}{\rm t}{\rm e}{\rm r}\;{\rm s}{\rm t}{\rm i}{\rm c}{\rm k}{\rm s}\;{\rm 
t}{\rm r}{\rm a}{\rm v}{\rm e}{\rm r}{\rm s}{\rm e}{\rm d}{\rm )}}}{{{\rm 
(}{\rm n}{\rm u}{\rm m}{\rm b}{\rm e}{\rm r}\;{\rm o}{\rm f}\;{\rm s}{\rm 
e}{\rm c}{\rm o}{\rm n}{\rm d}{\rm s}\;{\rm o}{\rm n}\;{\rm l}{\rm o}{\rm 
c}{\rm a}{\rm l}\;{\rm s}{\rm t}{\rm a}{\rm n}{\rm d}{\rm a}{\rm r}{\rm 
d}\;{\rm c}{\rm l}{\rm o}{\rm c}{\rm k}{\rm )}}}}{\rm .}
\end{equation}

Note that (\ref{eq28}) depends on the coordinate grid spacing, whereas (\ref{eq29}) does 
not. We therefore emphasize the following.

\textit{In general relativity the local coordinate speed of light varies with the spacetime coordinate grid chosen, but for spacetime frames whose line elements can be cast in form} (\ref{eq9})\textit{, the local physical (experimentally measured) speed of light is always c.}

Most spacetime frames treated in general relativity can have their line 
elements cast in the form of (\ref{eq9}). That is, they have no space-time cross 
terms and time is orthogonal to space. However, for frames that do have such 
cross terms\cite{Such:1}, care must be taken. For details, see 
Klauber\cite{Robert:1998}$^{{\rm ,}}$\cite{Robert:1}.

\section{SUMMARY}
\label{sec:summaryacetime}

Physical components of vectors (or tensors) are the component values one 
would measure by experiment with standard physical instruments. They equal 
the components associated with unit basis vectors. Within a given frame they 
are unique.

Coordinate components are those one uses in vector/tensor analysis for a 
particular coordinate system. They equal the component values associated 
with generalized basis vectors. They are not unique and vary with the 
coordinate grid chosen.

Physical components and coordinate components are related by (\ref{eq19}) for 
(contravariant) vectors and by (\ref{eq21}) for second order tensors.

In doing tensor analysis one must first convert known physical (measured) 
component values to coordinate components, then carry out the vector/tensor 
analysis, and as a final step convert the coordinate components answer to 
physical components for comparison with experiment. For invariant 
quantities, no conversion is necessary since they are the same for any 
coordinate system, they are not components of vectors or tensors, and they 
are already equivalent to physically measured values.

For time orthogonal frames:

1) the physical component in any direction equals the component value for 
the same direction found in a co-moving local Lorentz coordinate system,

2) the local physical speed of light is always $c$, and

3) the coordinate speed of light varies with the coordinate grid chosen, 
i.e., its numerical value is generally not equal to $c$.

\end{document}